\documentclass[a4paper,11pt]{article}
\usepackage{jinstpub} 
 \usepackage{subfigure}

\title{\boldmath The optimal structure of the MRPC detector for 0.511 MeV gamma based on Monte Carlo simulation}

 \author[a]{J. Liu,}
 \author[a]{Z. Chen,}
 \author[a,1]{Y. Wang,\note{Corresponding author.}}
 \author[a]{B. Guo,}
 \author[a]{D. Han}
 \author[a]{and Y. Li}
 \affiliation[a]{Key Laboratory of Particle and Radiation Imaging, Department of Engineering Physics, Tsinghua University, Beijing 100084, China}

\emailAdd{liu-jn20@mails.tsinghua.edu.cn}

\abstract{The detailed simulation of the Multi-gap Resistive Plate Chambers (MRPCs) provides the performance characteristics of MRPCs with different numbers of gas gaps and gap thicknesses. This helps in optimizing the structure of MRPCs under specific conditions by balancing time resolution, detection efficiency, and other performance metrics. To obtain the optimal structure of MRPCs for 0.511 MeV gammas, a complete simulation framework for gamma detection by the MRPCs based on Geant4 and Magboltz software is described in this paper. The simulation shows how gammas interact with MRPCs and the process of gas ionization, avalanche multiplication, and signal formation. The simulation results are in good agreement with the experimental results. By analyzing the time resolution and detection efficiency, the optimal structure of MRPCs for 0.511 MeV gammas is proposed.}


\begin{document}
\maketitle
\flushbottom

\section{Introduction}
\label{sec:intro}

The Multigap Resistive Plate Chamber (MRPC) is a spark-protected gas detector with good time resolution and low cost. In 1996, a team led by Williams proposed MRPC~\cite{zeballos1996new} for the first time based on the Resistive Plate Chamber (RPC). Essentially, the single millimeter-scale wide gas gap of RPC is divided into multiple narrow gas gaps with a width of hundreds of micrometers through glass electrodes. The narrow gas gap enables the Townsend avalanche to occur within a very small space, enhancing the avalanche's initial Townsend coefficient through intensified electric fields, ultimately refining time resolution without compromising detection efficiency. MRPC has excellent time resolution~\cite{ackermann2003star,akindinov2013performance} and good spatial resolution~\cite{shi2014high}, making MRPC Time-of-Flight Positron Emission Tomography (TOF-PET) a promising application prospect. The Fonte team demonstrated the feasibility of applying the RPC-PET method to specialized human brain PET scans. The initial results include the achievement of sub-millimeter image resolution, surpassing the state-of-the-art level, as well as the acquisition of detailed images of the striatal nuclei in a brain phantom \cite{fonte2023rpc}. Lippmann and Riegler discussed possible explanations for the differences in time resolution of Timing Resistive Plate Chambers for 0.511 MeV photons and particle beams, particularly focusing on the statistical fluctuations of deposited charge and the flight time distribution of Compton electrons \cite{lippmann2009simulation}.   

Currently, international research on the simulation of gamma detection by the MRPC mainly focuses on studying the relationship between the number of gas gaps, the thickness of the converter, and the detection efficiency\cite{nizam2019development,weizheng2014monte}. However, the time resolution is of utmost importance for MRPC TOF-PET. Therefore, in this paper, we have established a comprehensive simulation framework for detecting gamma rays by MRPCs. This framework encompasses the entire simulation process, from detector construction and particle generator setup to energy deposition modeling, avalanche multiplication calculation, induced signal generation, waveform shaping, and ultimately, data storage and analysis. Using this framework, we have conducted detailed studies on the time resolution and detection efficiency of various detector structures, ultimately proposing an optimized structure optimized for 0.511 MeV gamma rays.

\section{Simulation framework for gamma detection by MRPC}

\begin{figure}[htbp]
\centering
\includegraphics[width=0.5\textwidth]{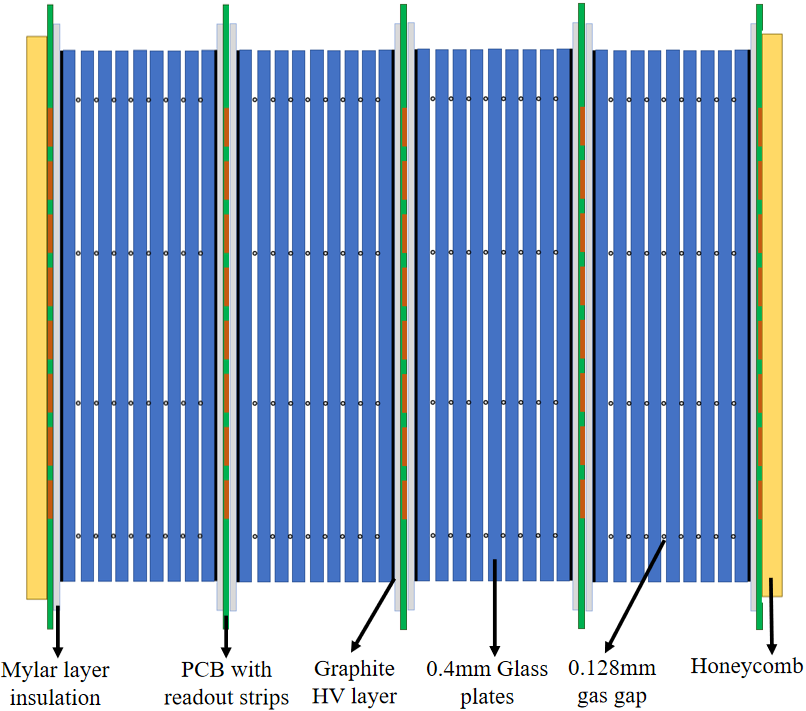}
\caption{Schematic drawing of the cross-section of the 4-stack 32-gap MRPC detector.\label{fig:1}}
\end{figure}

The MRPC detector is modeled in Geant4, including the geometry and material configuration. The schematic drawing of the cross-section of the 4-stack 32-gap MRPC is shown in figure~\ref{fig:1}. Nylon fishing lines create uniform gaps between the glass plates. Mylar layers are used to separate the graphite high-voltage layer from the printed circuit board (PCB). Honeycomb panels are used to support and protect the detector. In the simulation, parameters such as the number of stacks and gaps, the thickness of the resistive plates and gas gaps can be changed as needed. The source of particles can be defined as any type, with any energy, placed at any position, and emitted in any direction. 

\begin{figure}[htbp]
\centering
\subfigure[]{
\label{fig：subfig_a}
\includegraphics[width=.45\textwidth]{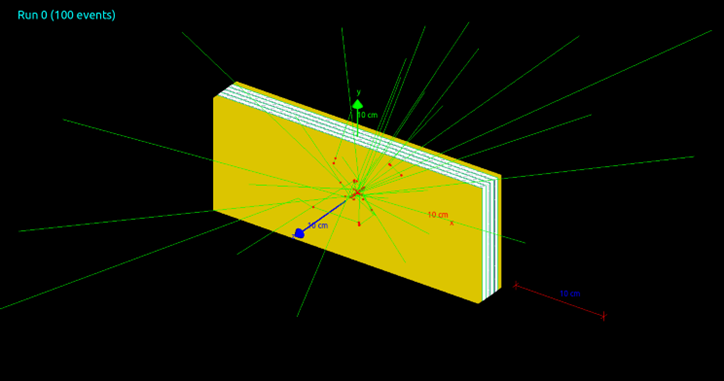}
}
\subfigure[]{
\label{fig：subfig_b}
\includegraphics[width=.45\textwidth]{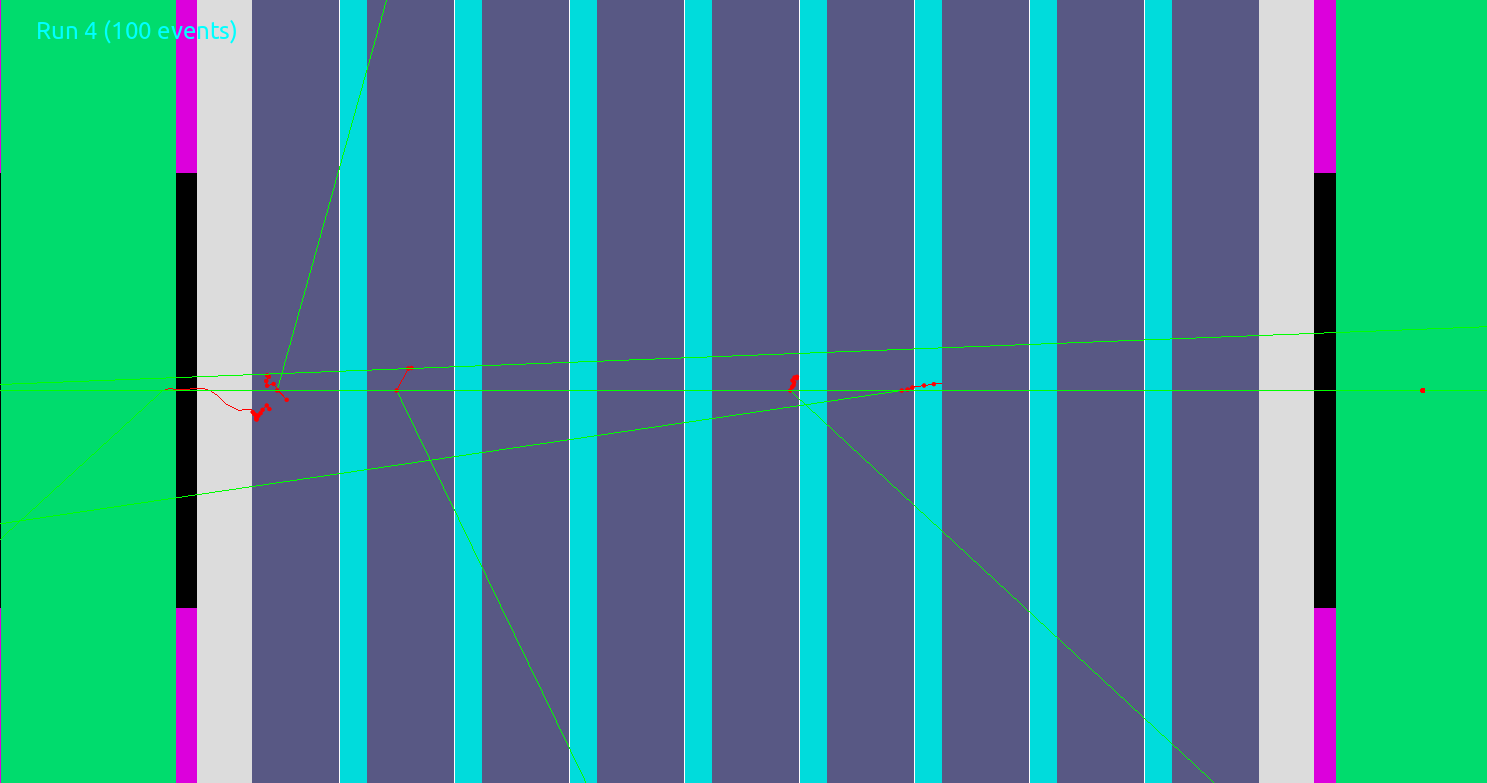}
}
\caption{Visualization interface in Geant4 simulation. (a) overall, (b) local amplification of one stack.\label{fig:2}}
\end{figure}

In this paper, the particle source is defined as 0.511 MeV gamma, incident in the direction perpendicular to the detector. Figure~\ref{fig:2} shows the visual interface in the Geant4 simulation. 

\begin{figure}[htbp]
\centering
\includegraphics[width=0.5\textwidth]{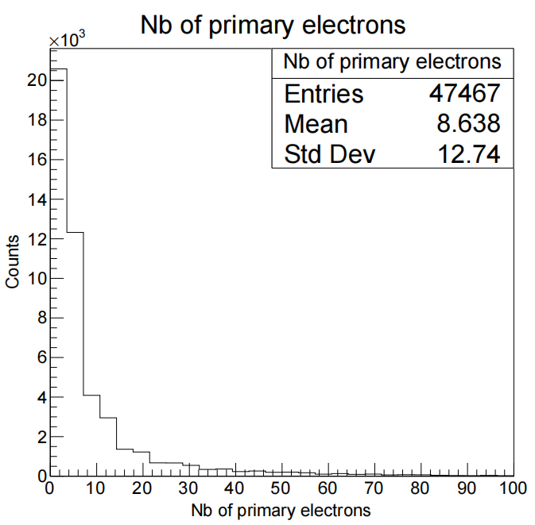}
\caption{The number of primary electrons in the simulation for the 4-stack 32-gap MRPC with 0.128 mm gap thickness.\label{fig:10}}
\end{figure}

The green tracks are gamma photons and the red tracks are electrons. The incident gamma photons interact with the glass plates and produce electrons. Some of these electrons deposit their energy within the glass plates, while others enter the gas gaps. The electrons that enter the gas gaps drift along the field lines toward the anode, ionizing gas molecules as they move. Each primary ionization leads to an avalanche. Taking the 4-stack 32-gap MRPC with 0.128 mm gap thickness, as an example, the distribution of the primary electron numbers is shown in figure \ref{fig:10}.

\begin{figure}[htbp]
\centering
\subfigure[]{
\label{fig：subfig_c}
    \includegraphics[width=.45\textwidth]{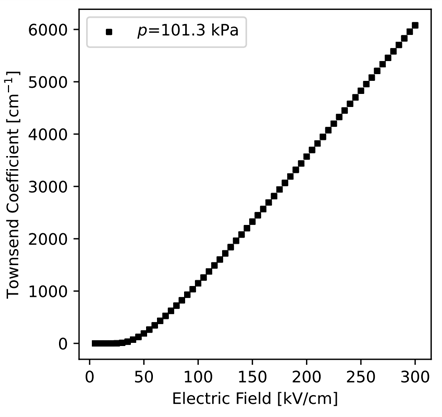}
}
\subfigure[]{
\label{fig：subfig_d}
\includegraphics[width=.45\textwidth]{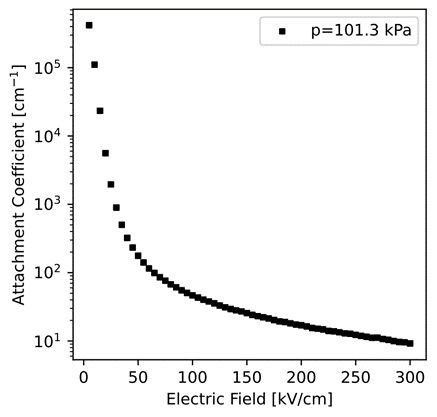}
}
\subfigure[]{
\label{fig：subfig_e}
    \includegraphics[width=.45\textwidth]{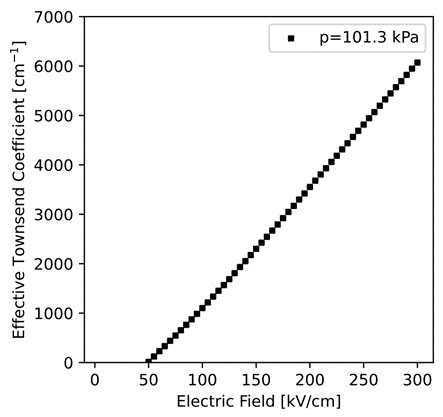}
}
\subfigure[]{
\label{fig：subfig_f}
    \includegraphics[width=.45\textwidth]{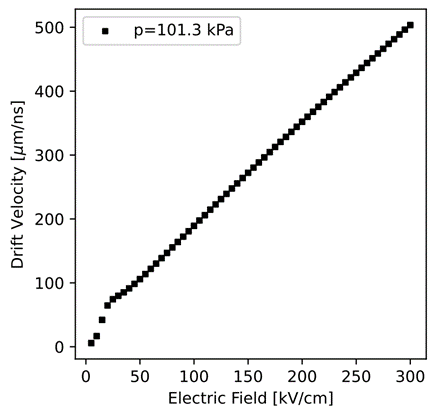}
}
\caption{(a) Townsend coefficient, (b) attachment coefficient, (c) effective Townsend coefficient and (d) drift velocity calculated by Magboltz.\label{fig:3}}
\end{figure}

In the LHC simulation framework, energy deposition is typically simulated using the EMstandard physics process list within the Geant4 package \cite{allison2016recent}. However, this model does not account for shell electron effects and is mainly suitable for thick sensors \cite{wang2018impact}. Geant4 also offers a more detailed energy loss model, the Photo Absorption Ionization (PAI) model \cite{apostolakis2000implementation}. This model relies on an adjusted table of photo-absorption cross-section coefficients and is compatible with various elements. It believes that the interaction between charged particles and extranuclear electrons is accomplished by emitting and absorbing virtual photons, the range of virtual photons is the reaction cross section, so the ionization energy loss can be calculated by using the photo-absorption data of ultraviolet and X-ray interactions with matter. Experimental data demonstrate that this model's energy loss predictions align well with results for thin sensors \cite{allison1980relativistic}. Because the gas gap thickness of MRPC detector is generally in the order of several hundred microns, the PAI model in Geant4 is chosen for energy deposition in thin sensors. Electrons from the Gamma photon interacting with the MRPC detector drift towards the anode and start the avalanche multiplication. The Avalanche follows the Townsend multiplication law \cite{townsend1900conductivity}. The change of the average number of electrons in the gas gap with the drift distance is \cite{riegler2003detector}: 
\begin{equation}
\label{eq:1}
\begin{aligned}
\frac{{d\bar n}}{{dx}} = \left( {\alpha  - \eta } \right)\bar n
\end{aligned}
\end{equation}
where $ \alpha $ is the Townsend coefficient, $ \eta $ is the attachment coefficient and $\bar n$ is the number of electrons at position $x$. The Townsend coefficient, the attachment coefficient, the effective Townsend coefficient, and the drift velocity are calculated by Magboltz, as shown in figure \ref{fig:3}.

The induced current on the readout electrodes follows the Ramo theory \cite{ramo1939currents}, see \eqref{eq:2} \cite{riegler2003detector,raether1964electron}.
\begin{equation}
\label{eq:2}
\begin{aligned}
i(t) = \frac{{{E_w} \cdot v}}{{{V_w}}}{e_0}N(t),
\qquad
\frac{{{E_w}}}{{{V_w}}} = \frac{\varepsilon }{{ng\varepsilon  + (n + 1)d}}
\end{aligned}
\end{equation}
where $v$ is the drift velocity of electrons, ${e_0}$ is the electron charge, $N(t)$ is the number of electrons in the gas gap at time $t$, and ${E_w}$ is the weighting field, whose value is equal to the electric field when the potential of the readout electrode is set to ${V_w}$ and others 0. The value of $\frac{{{E_w}}}{{{V_w}}}$ is related to the geometry of the detector and the material of the resistive plate. $n$ is the number of gas gaps, $g$ and $d$ are the thickness of the gas gap and the resistive plate respectively, $\varepsilon $ is the relative dielectric constant of the resistive plate.

\begin{figure}[htbp]
\centering
\subfigure[]{
\label{fig：subfig_g}
\includegraphics[width=.45\textwidth]{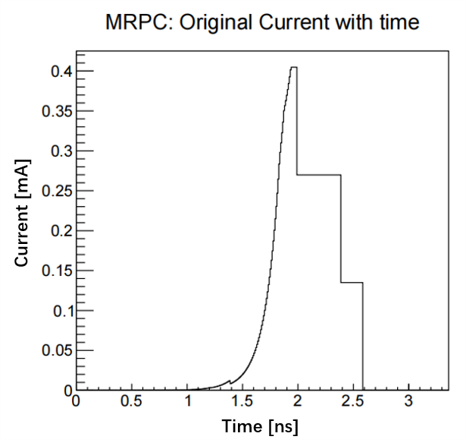}
}
\subfigure[]{
\label{fig：subfig_h}
\includegraphics[width=.42\textwidth]{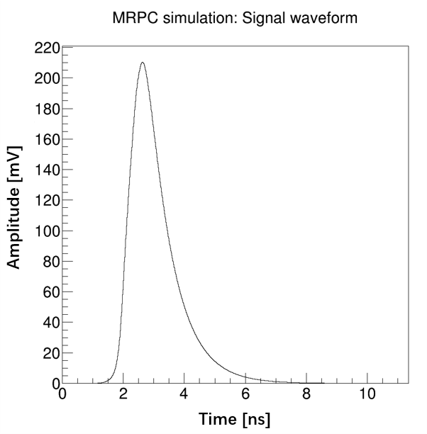}
}
\caption{(a) Original induced current, (b) The signal convoluted with readout electronics response.\label{fig:4}}
\end{figure}

\begin{figure}[htbp]
\centering
\subfigure[]{
\label{fig：subfig_g}
\includegraphics[width=.44\textwidth]{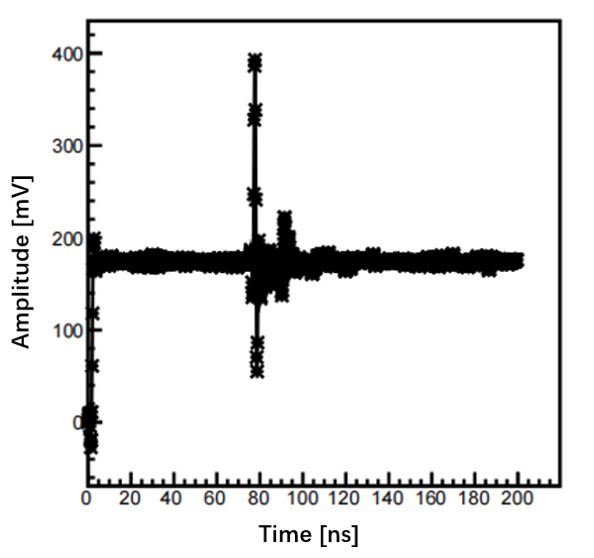}
}
\subfigure[]{
\label{fig：subfig_h}
\includegraphics[width=.45\textwidth]{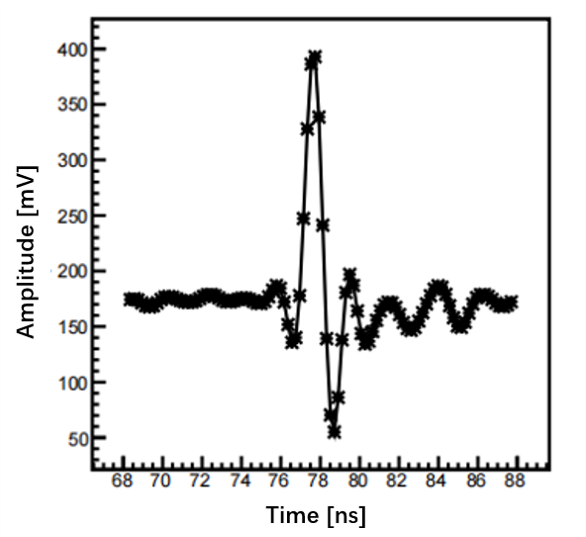}
}
\caption{(a) A complete waveform of the measured signal in the experiment, (b)  local amplification of the waveform.\label{fig:11}}
\end{figure}

Taking the 4-gap MRPC with 0.4 mm gap thickness as an example, the change of induced current with time is shown in figure \ref{fig:4}. In the simulation, the signal can be represented as the Fourier convolution of the original induced current $i(t)$ and a simplified electronics response $f(t)$.
\begin{equation}
\label{eq:3}
\begin{aligned}
V(t) = i(t) \otimes f(t),
\qquad
f(t) = A\left( {{e^{ - \frac{t}{{{\tau _1}}}}}} \right. - \left. {{e^{ - \frac{t}{{{\tau _2}}}}}} \right)
\end{aligned}
\end{equation}

where $A$ is the amplification factor of the electronics, and ${\tau _1}$ and ${\tau _2}$ correspond to the time constants of the ${\rm{RC}}$ circuits in the electronics. ${\tau _1}$ and ${\tau _2}$ affect the length of the leading and trailing edge of the output signal, and thus these values should be adjusted according to the electronics used in the experiment. Figure \ref{fig:4}(b) is the same example as figure \ref{fig:4}(a), and its shape is very similar to the real waveform. Figure \ref{fig:11} shows a waveform of the measured signal in the experiment. In the experiment, the electronics and all the cables will bring noise to the MRPC signal. In the simulation, a Gaussian noise is used to simulate it, and the $\sigma $ of the noise is consistent with the experiment.

\section{Results of the simulation and comparison with experimental data}
\begin{figure}[htbp]
\centering
\subfigure[]{
\label{fig：subfig_k}
    \includegraphics[width=.45\textwidth]{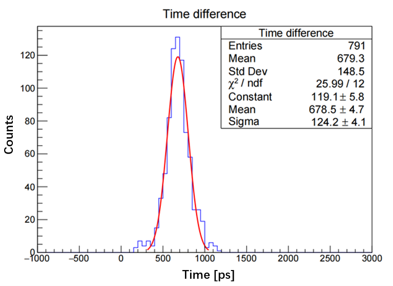}
}
\subfigure[]{
\label{fig：subfig_l}
    \includegraphics[width=.45\textwidth]{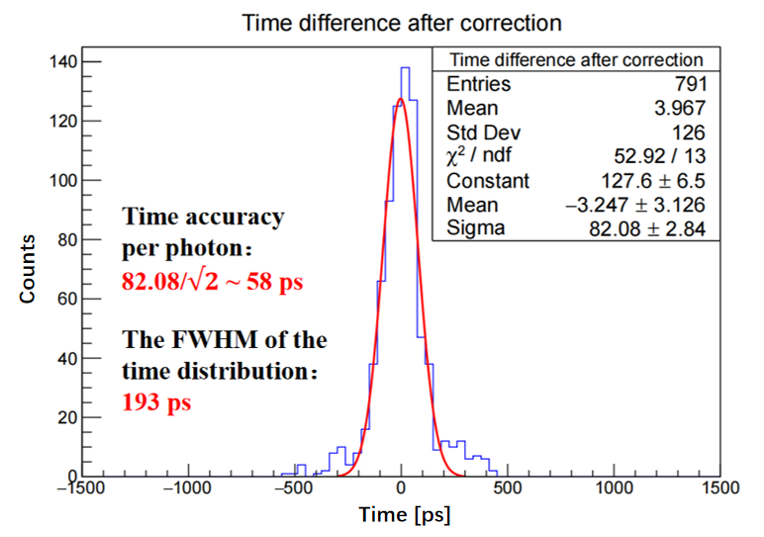}
}
\caption{The distribution of time difference before (a) and after (b) time walk correction obtained by the experiment of 4-stack 32-gap MRPCs.\label{fig:5}}
\end{figure}
\begin{figure}[htbp]
\centering
\subfigure[]{
\label{fig：subfig_m}
    \includegraphics[width=.45\textwidth]{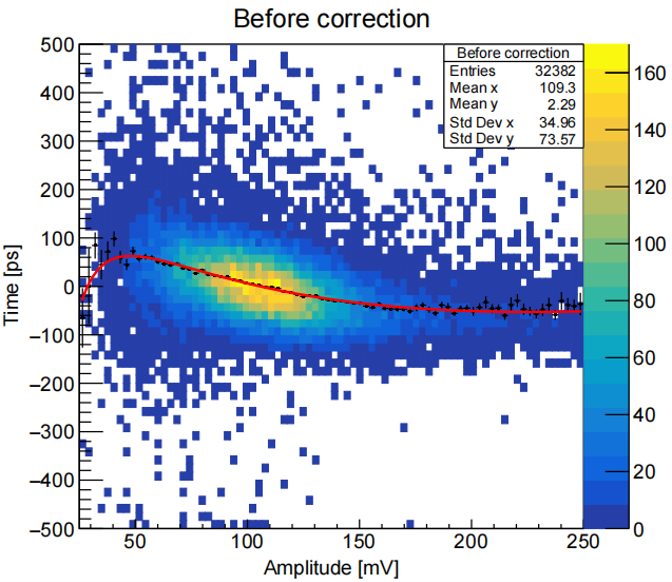}
}
\subfigure[]{
\label{fig：subfig_n}
\includegraphics[width=.45\textwidth]{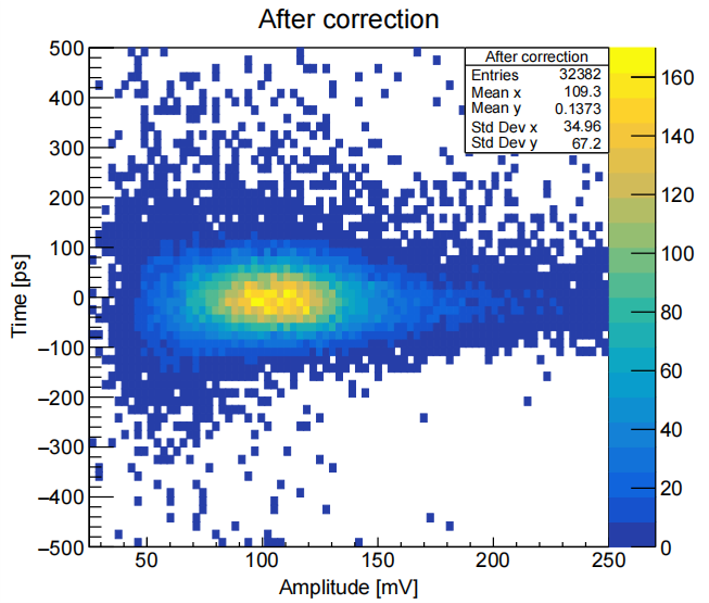}
}
\subfigure[]{
\label{fig：subfig_o}
    \includegraphics[width=.45\textwidth]{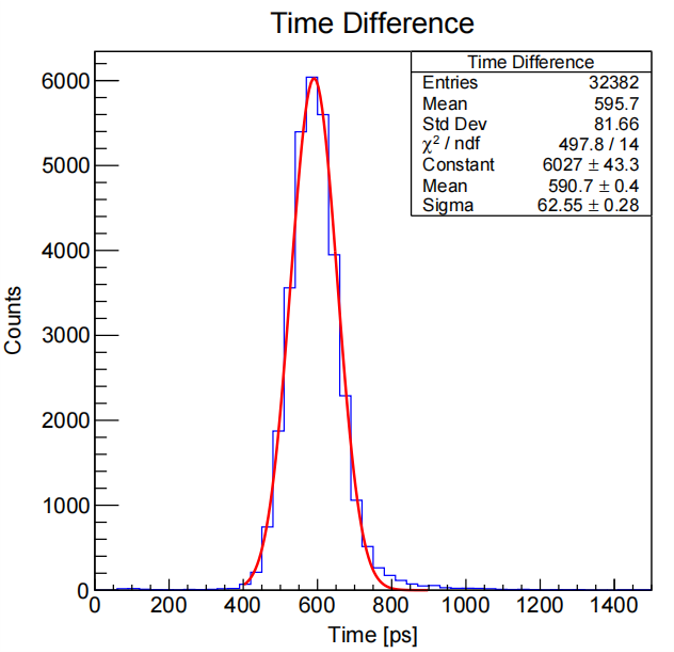}
}
\subfigure[]{
\label{fig：subfig_p}
    \includegraphics[width=.45\textwidth]{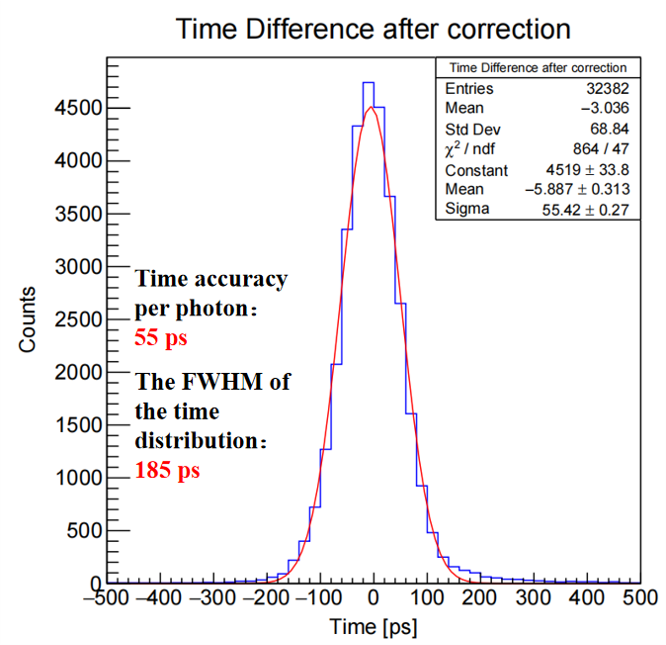}
}
\caption{The relationship between time difference and signal amplitude before (a) and after (b) time walk correction, and the distribution of time difference before (c) and after (d) time walk correction obtained by the simulation of 4-stack 32-gap MRPCs.\label{fig:6}}
\end{figure}

To evaluate the effectiveness of the simulation, the simulation and experimental data of two different MRPC detectors with two kinds of readout systems are compared. Two 4-stack 32-gap MRPC prototypes with 0.128 mm gap thickness are fabricated and tested. Figure \ref{fig:5} shows the distribution of time difference before and after time walk correction obtained by the experiment of 4-stack 32-gap MRPCs. The time accuracy per photon is 58 ps. The obtained FWHM of the time distribution for 0.511 MeV photons is 193 ps by using the fast front-end amplifier and the Tektronix oscilloscope. The simulation conditions correspond to the experiment. The relationship between time difference and signal amplitude before and after time walk correction is shown in figure \ref{fig:6}(a)(b). Figure \ref{fig:6}(c)(d) shows the distribution of time difference before and after time walk correction obtained by the simulation of 4-stack 32-gap MRPCs. The time accuracy per photon is 55 ps. The simulated FWHM of the time distribution for 0.511 MeV photons is 185 ps, which is consistent with the experimental results.

In addition, the simulation and experimental results of 4-gap MRPC with 0.4 mm gap thickness are compared. The simulated FWHM of the time distribution for 0.511 MeV photons is 616 ps, which is consistent with the experimental results of 665 ps FWHM measured by the fast front-end amplifier and waveform digitizer system.

\section{The optimal structure of MRPC for 0.511 MeV gamma}
\begin{figure}[htbp]
\centering
\subfigure[]{
\label{fig：subfig_q}
\includegraphics[width=.45\textwidth]{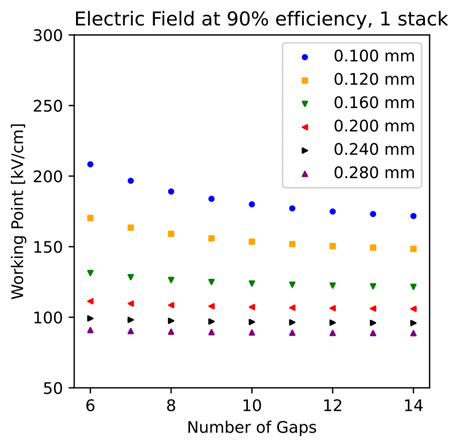}
}
\subfigure[]{
\label{fig：subfig_r}
\includegraphics[width=.45\textwidth]{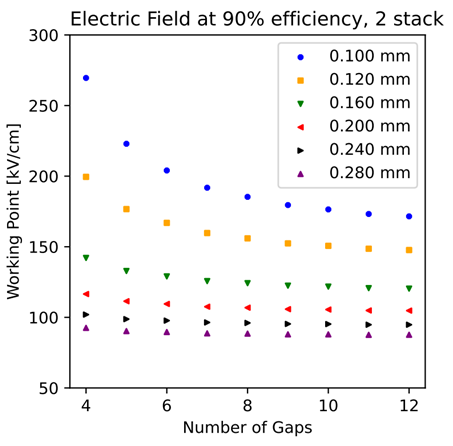}
}
\caption{The working electric field of MRPC detectors.\label{fig:7}}
\end{figure}

\begin{figure}[htbp]
\centering
\subfigure[]{
\label{fig：subfig_s}
\includegraphics[width=.45\textwidth]{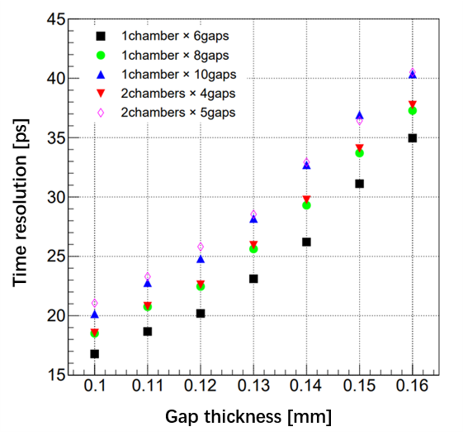}
}
\subfigure[]{
\label{fig：subfig_t}
\includegraphics[width=.45\textwidth]{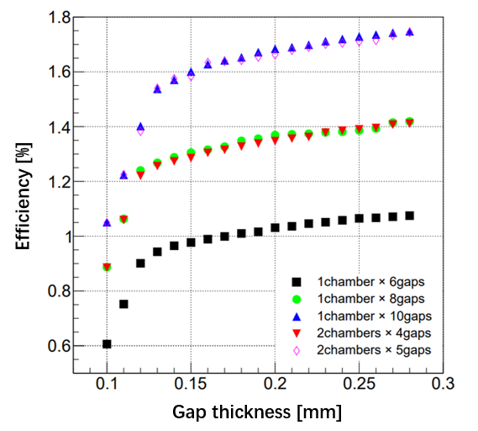}
}
\caption{The intrinsic time resolution (a) and
detection efficiency (b) of MRPC with different structures.\label{fig:8}}
\end{figure}

The working electric field is defined as the corresponding electric field when the detection efficiency of MRPC detector for cosmic ray reaches more than 90\%. The working electric field of MRPC detectors with different structures is calculated by Magboltz software, as shown in figure \ref{fig:7}. As can be seen from the figure  \ref{fig:7}, the thinner the thickness of the gas gap, the higher the working electric field. The intrinsic time resolution and detection efficiency of MRPC with different structures are simulated as shown in figure \ref{fig:8}. As the gas gap thickness and the number of gaps decreases, the intrinsic time resolution improves significantly, albeit at the cost of reduced detection efficiency. Specifically, when the gas gap thickness surpasses 0.13 mm, the detection efficiency plummets due to the inadequate signal amplitude generated. Therefore, balancing the need for exceptional time resolution with adequate detection efficiency, the optimal detector configuration comprises a single stack with eight gas gaps, each measuring 0.13 mm in thickness. This design ensures an optimal trade-off between performance metrics. 

The relationship between the detection efficiency, the induced charge of the MRPC detector and the thickness of the resistive plate is simulated. The results reveal an intriguing trend: as the thickness of the resistive plate increases, the detection efficiency initially exhibits an upward surge, followed by a gradual decline. The induced charge consistently decreases with the augmentation of the resistive plate's thickness. Based on these findings, the optimal thickness for the resistive plate is determined to be 0.4 mm.

\begin{figure}[htbp]
\centering
\includegraphics[width=0.45\textwidth]{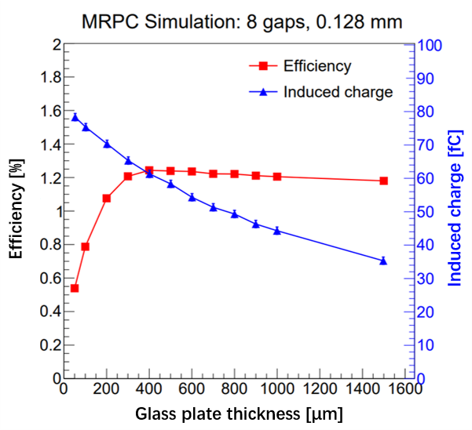}
\caption{Results of detection efficiency and induced charge of MRPC detector with different resistive plate thickness.\label{fig:9}}
\end{figure}
\section{Conclusion}
We have developed a comprehensive simulation framework for the gamma detection by MRPCs, and it has been confirmed that the simulation results are in good agreement with the experimental results obtained by using a ${}^{22}{\rm{Na}}$ radioactive source. This simulation framework can be utilized to optimize the detector design. For 0.511 MeV gammas, balancing time resolution and detection efficiency, the optimal detector structure is determined to be 1-stack 8-gap MRPC with 0.13 mm gas gap thickness and 0.4 mm resistive plate thickness. Additionally, the simulated complete signal waveforms provide a foundation for further improving the time resolution through the reconstruction of time using algorithms such as machine learning.

\acknowledgments

The work is supported by the National Natural Science Foundation of China under Grant No. 11927901, 11420101004, 11461141011, 11275108, 11735009 and U1832118. This work is also supported by the Ministry of Science and Technology under Grant No. 2020YFE0202001, 2018YFE0205200 and 2016YFA0400100.


 \bibliographystyle{JHEP}
 \bibliography{reference}

\end{document}